\documentstyle[12pt]{crnart12}
\def\NP{{\it Nucl.Phys. }}
\def\PL{{\it Phys.Lett. }}
\def\MPL{{\it Mod.Phys.Lett. }}
\def\SPJ{{\it Sov.J.Phys. }}
\def\PR{{\it Phys.Rev. }}
\def\AM{{\it Adv.Math. }}
\def\IJMP{{\it Int.J.Mod.Phys. }}
\def\be{\begin{equation}}
\def\ee{\end{equation}}
\def\bea{\begin{eqnarray}}
\def\eea{\end{eqnarray}}
\def\bq{\begin{quote}}
\def\eq{\end{quote}}
\def\bseq{\begin{subequation}}
\def\eseq{\end{subequation}}
\def\bsea{\begin{subeqnarray}}
\def\esea{\end{subeqnarray}}
\def\simlt{\mathrel{\lower2.5pt\vbox{\lineskip=0pt\baselineskip=0pt
           \hbox{$<$}\hbox{$\sim$}}}}
\def\simgt{\mathrel{\lower2.5pt\vbox{\lineskip=0pt\baselineskip=0pt
           \hbox{$>$}\hbox{$\sim$}}}}

\begin{document}
\pagestyle{empty}
\begin{flushright}
CERN-TH.6799/93
\end{flushright}
\vglue 2.5cm
\begin{center}
{\bf FOUR-DIMENSIONAL GRAVITATIONAL BACKGROUNDS}\\
{\bf  BASED ON
 $N=4$,~$\hat{c}=4$, SUPERCONFORMAL SYSTEMS }\\
\vglue 1.5cm
{\bf Costas Kounnas$^{*)}$}
\\
\vglue.3cm
Theoretical Physics Division, CERN \\
Geneva, Switzerland \\
\vskip 1cm
{\bf ABSTRACT}
\end{center}

\noindent
We propose two new realizations of the $N=4$,
$\hat{c}=4$ superconformal system
 based on the compact and non-compact versions
 of parafermionic algebras. The
 target space interpretation of these systems is given in terms of
 four-dimensional target spaces with non-trivial metric
 and topology different from
 the previously known four-dimensional semi-wormhole realization.
 The proposed
 $\hat{c}=4$ systems can be used as a
 building block to construct perturbatively
 stable superstring solutions with
 covariantized target space supersymmetry
 around non-trivial gravitational and dilaton backgrounds.
\vspace*{5cm}

\noindent
\rule[.1in]{16.5cm}{.002in}

\noindent
$^{*)}$ On leave from Ecole Normale Sup\'erieure, 24, rue Lhomond, 75231 Paris
Cedex  05, France.
\vspace*{0.5cm}

\begin{flushleft}
CERN-TH.6799/93\\
hep-th/9304102\\
February 1993
\end{flushleft}
\vfill\eject

\setcounter{page}{1}
\pagestyle{plain}

\noindent{\bf 1. INTRODUCTION}

The study of stable string solutions around non-trivial gravitational
 backgrounds provides a better understanding of the quantum gravitational
 phenomena; in principle,
 it may also shed some light on some physical processes
 in the presence of a highly curved or even singular space-time (e.g.
 black-holes, wormholes, singular cosmological solutions or other)
 \cite{all}--\cite{hor}.

Approximate classical or semi-classical solutions can be obtained via the
 $\alpha'$ expansion of the two-dimensional $\sigma$-model
 \cite{aaa}. These solutions (in
 lowest order of $\alpha'$) are identical to
 those of the classical Einstein
 equations in the presence of a dilaton field,
 antisymmetric tensor field and
 some  minimally coupled gauge matter.
 The $\alpha'\quad \sigma$-model approach is
 very useful since it provides us with a well-defined
 method to obtain and study, in
 low energies, the string-induced effective field theories. The $\alpha'$
 expansion however breaks down and it fails
 to describe the interesting physics
 of the quantum gravitational phenomena in
 short distances, in particular all
 stringy phenomena around highly curved or
 singular backgrounds. It is therefore
 necessary to go beyond the $\alpha'$ expansion
 and try to extend the physically
 interesting (semi)-classical solutions to some
 {\em exact string vacua}.

In our days we do not know yet of any systematic
way to extend the $\sigma$-model
 classical solutions to some exact
 (super)-conformal theories. This extension is
 generally difficult and not obvious at all, since
 it requests the existence of
 an exactly solvable (super)-conformal theory
 defined on the two-dimensional world-sheet.
 Unfortunately, the number of the known (super)-conformal systems is very
 limited and even more restricted for the
 systems with some interesting and non-trivial target-space interpretation.

 One way to overcome this difficulty is by
 constructing directly non-trivial
 string solutions, via the conformal block-construction method,
 by tensoring known
 conformal systems. Then one tries to interpret
 the obtained string vacua in terms
 of the target-space backgrounds. In the conformal block construction, the
 $\alpha'$ expansion is not necessary and is used
 just as a suitable target-space interpretation.

In this work we will construct two new families
of four-dimensional target
 spaces. The first one is denoted by
 $\Delta^{(4)}_{k}$,  the
 second one by $C^{(4)}_{k} (k=1,2,...$).
 The  $\Delta^{(4)}_{k}$ space  is
 defined as a direct product of two superconformal models, e.g. two
 supersymmetric gauged WZW models:

$$
\Delta^{(4)}_{k}=
\left(\frac{SU(2)}{U(1)}\right) ^{ss}_{k}
\bigotimes
\left(\frac{SL(2,R)}{U(1)}\right)^{ss}_{k'}
$$
with,
$$
\ k'=k+4
\, .
\eqno(1)
$$

The central charge of $\Delta^{(4)}_{k}$
is always $c=6 ~~(\hat{c}=4$) for any
 value of $k$, thanks to the relation among the
 $SU(2)$ and $SL(2,R)$ levels. The
 central-charge deficit coming from the
 {\em negatively} curved $SU(2)/U(1)$ two-dimensional sub-space, it is
cancelled
by
 the central charge benefit coming from
 the {\em positively} curved $SL(2,R)/U(1)$ two-dimensional sub-space.
 The value of the central
 charge ($c=6$, or $\bf \hat{c}$ =4)  is the same with that of the
four-dimensional flat space $F^{(4)}$, with four free super-coordinates.

The $C^{(4)}_{k}$ space  is also a product of
two superconformal models:
$$
C^{(4)}_{k}=\left(\frac{SU(2)}{U(1)}\right) ^{ss}_{k}
\bigotimes
\left(U(1)_{R}\bigotimes U(1)_{Q}
\right)^{ss}_{k}
$$
with
$$
 Q=\sqrt{\frac{2}{k+ 2}}~~~~~~~~~~~~~~{\rm and}~~~~~~~~~~~~~~R=\sqrt{2k}
\, .
\eqno(2)
$$

The $C^{(4)}_{k}$ has also fixed central charge,
$c=6$ ($\hat{c}=4$) for any
 value of $k$ as in $\Delta^{(4)}_{k}$.
 Here the central-charge deficit coming
 from the $SU(2)/U(1)$ two-dimensional sub-space
 is balanced, owing to the relation
 among the $SU(2)$ level $k$ and the background
 charge $Q$ on  one of the two
 Abelian coordinates.
 The other coordinate is flat but compactified on a torus
 of radius $R$, which is also  fixed  by $k$ [see eq. (2)].

What we would like to show is that the
$\Delta^{(4)}_{k}$ and $C^{(4)}_{k}$
 spaces are  very special deformations of the $F^{(4)}$ flat space. These
 deformations are parametrized in $\Delta^{(4)}_{k}$
 by the levels $k$ and $k'$.
 For $k'=k+4$ the $N=4$ superconformal properties of
 the flat space  remain
 valid in the deformed (curved) space $\Delta^{(4)}_{k}$.
In $C^{(4)}_{k}$ the deformation is given by the level
$k$, the background charge
 $Q$, and the radius $R$. For the special relations among
 $k$, $Q$ and $R$ as
 given in eq. (2), $C^{(4)}_{k}$ has the $N=4$
 superconformal properties of the
 flat space.

In order to make the similarities and differences of the
 $F^{(4)}$, $\Delta^{(4)}_{k}$ and $C^{(4)}_{k}$
 spaces more transparent, we will first present
 the realization  of the  $N=4, {\bf \hat{c}}=4$, in terms of the free
 super-coordinates, and then  present the modifications that are
 necessary to define the new realizations in the deformed spaces.

\vspace*{1cm}
\noindent
{\bf 2. $N = 4$ SUPERCONFORMAL ALGEBRA BASED ON $F^{(4)}, \Delta^{(4)}_K$ and
$W^{(4)}_k$}
\vspace*{0.3cm}
\noindent
{\bf 2.1. $F^{(4)}$ realization. Free supercoordinates}
\cite{ff},\cite{ggg}

The basic operators of the $N=4$ algebra,
(i) the stress tensor $T_{B}$, (ii) the
 four supercurrents $G_{i}$, and (iii) the $SU(2)_{n}$ level $n=1$ currents
 $S_{i}$, are constructed in terms of the four free
 super-coordinates, ($J_{i}, \Psi_{j}, i,j=1,2,3,4$):
$$
J_{i}(\xi)J_{j}(\xi')\equiv -\frac{\delta_{ij}}{(\xi-\xi')^{2}}
$$
$$
\Psi_ {i}(\xi)\Psi_{j}(\xi')\equiv-\frac{\delta_{ij}}{(\xi-\xi')}
\, .
\eqno(3)
$$

The (right-moving) basic operators are given as follows:
$$
T_{B}=-\frac{1}{2}\left(J^{2}_{i}-\Psi_{i}\partial\Psi_{i}\right)
$$
\break\newpage

$$
G_{1}=+J_{1}\Psi_{1}+J_{2}\Psi_{2}+J_{3}\Psi_{3}+J_{4}\Psi_{4}
$$
$$
G_{2}=+J_{1}\Psi_{2}-J_{2}\Psi_{1}-J_{3}\Psi_{4}+J_{4}\Psi_{3}
$$
$$
G_{3}=-J_{1}\Psi_{4}+J_{2}\Psi_{3}- J_{3}\Psi_{2}+J_{4}\Psi_{1}
$$
$$
G_{4}=-J_{1}\Psi_{3}-J_{2}\Psi_{4}+J_{3}\Psi_{1}+J_{4}\Psi_{2}
$$

$$
S_{i}=\frac{1}{2}\left(\Psi_{4}\Psi_{i}+
\epsilon_{ijl}\Psi_{j}\Psi_{l}\right)
\, .
\eqno(4)
$$

Using the free-field Operator Product Expansion
(OPE), one finds  the
 relations of the $N=4$ algebra with ${\bf \hat{c}}=4$:
$$
G_{4}(\xi)G_{4}(\xi')\equiv
\frac{\bf\hat{c}}{(\xi-\xi')^{3}}+\frac{2T_{B}(\xi')}
 {(\xi-\xi')}
$$
$$
G_{i}(\xi)G_{j}(\xi')\equiv
 \delta_{ij}\frac{\bf\hat{c}}{(\xi-\xi')^3}-4\epsilon_{ijl}
 \frac{S_{l}(\xi')}{(\xi-\xi')^2}+
 2\delta_{ij}\frac{T_{B}(\xi)}{(\xi-\xi')}
$$
$$
G_{4}(\xi)G_{i}(\xi')\equiv\frac{4S_{i}(\xi')}{(\xi-\xi')}
$$
$$
S_{i}(\xi)G_{4}(\xi')\equiv-\frac{G_{i}(\xi')}{2(\xi-\xi')}
$$
$$
S_{i}(\xi)G_{j}(\xi')\equiv\frac{1}{2(\xi-\xi')}
\left(\delta_{ij}G_{4}(\xi')
 +\epsilon_{ijl}G_{l}(\xi')\right)
$$
$$
S_{i}(\xi)S_{j}(\xi')\equiv-\delta_{ij}\frac{n}{2(\xi-\xi')^2}
 +\epsilon_{ijl}\frac{S_{l}(\xi')}{(\xi-\xi')}
\, .
\eqno(5)
$$
In the above equations the level $n$ of the $SU(2)$
currents and the central charge
 $\hat{\bf {c}}$ are related by $\hat{\bf {c}}=4n$, \cite{ff},\cite{ggg};
 therefore the level $n$ is fixed to one $(n=1)$ for $\hat{\bf {c}}=4$.

For later convenience it is useful to adopt a complex notation for the
 coordinate currents,
$$
P=J_{1}+iJ_{2} \, , \;\;\;\;\;  P^{t}=-J_{1}+iJ_{2}~,
$$
$$
\Pi=J_{4}+iJ_{3} \, , \;\;\;\;\;  \Pi^{t}=-J_{4}+iJ_{3}~,
$$
with
$$
P(\xi)P^{t}(\xi')\equiv\frac{2}{(\xi-\xi')^{2}}+2T_{P}
$$
$$
\Pi(\xi)\Pi^{t}(\xi')\equiv\frac{2}{(\xi-\xi')^{2}}+2T_{\Pi}
\, ,
\eqno(6)
$$
where $T_{P}$ and $T_{\Pi}$ are the stress tensor of the ($P,P^{t}$)
 and ($\Pi,\Pi^{t}$) conformal sub-systems.

It is also  useful to bosonize the free fermions
in terms of two bosons, $H^{+}$
 and $H^{-}$. The desired bosonization must express
 the global properties of the
 $SO(4)$ level one fermionic currents, $\Psi_{i}\Psi_{j}$.

 First, we decompose the $SO(4)_{1}$ currents in
 terms of two $SU(2)$ level-one
 currents using the {\em {self-dual}} and
 {\em {anti-self-dual}} projections,
$$
S_{i}=\frac{1}{2}\left(+\Psi_{4}\Psi_{i}+
\frac{1}{2}\epsilon_{ijl}\Psi_{j}
 \Psi_{l}\right)
$$
$$
\tilde{S}_{i}=\frac{1}{2}\left(-\Psi_{4}\Psi_{i}+
\frac{1}{2}\epsilon_{ijl}\Psi_{
 j} \Psi_{l}\right)
\, .
\eqno(7)
$$

Then, we parametrize $S_{i}$ and $\tilde{S}_{i}$ in terms of the two free
 bosons, $H^{+}$ and $H^{-}$, both of them
 compactified on the torus with self-dual
 radius [the $SU(2)$-extended symmetry points]:
$$
R_{H^{+}}=R_{H^{-}}= \sqrt{2}~,
$$
giving
$$
(S_{i})= \left( \frac{1}{2} \partial H^{+} \, , \;\;
e^{ \pm i \sqrt{2} H^{+}}
 \right)
\, , \;\;\;\;\;\;\;\;\; SU(2)_{H^{+}},
$$
$$
(\tilde{S}_{i})= \left( \frac{1}{2}
\partial H^{-} \, , \; \; e^{ \pm i \sqrt{2}
 H^{-}} \right) \, , \;\;\;\;\;\;\;\;\; SU(2)_{H^{-}}
\, .
\eqno(8)
$$

In terms of $P,P^{t}$, $ \Pi ,\Pi^{t}$,
$H^{+}$ and $H^{-}$ the basic operators
 of the $N=4$ algebra [see eq. (4)] become:
$$
T_{B}= - \frac{1}{2} \left((\partial H^{+})^{2}+
(\partial H^{-})^{2}-PP^{t}
 -\Pi \Pi^{t} \right)
$$
$$
{\bf {G}} = -\left( \Pi^{t}e^{- \frac{i}{\sqrt{2}}H^{-}} + P^{t}e^{+ \frac{i}{
 \sqrt{2}} H^{-}} \right) e^{+ \frac{i}{ \sqrt{2}}H^{+}}
$$
$$
{\bf {\tilde {G}}}= \left( \Pi \;\; e^{+ \frac{i}{\sqrt{2}}H^{-}} - P \;\; e^{-
 \frac{i}{ \sqrt{2}} H^{-}} \right) e^{+ \frac{i}{ \sqrt{2}}H^{+}}
$$
$$
(S_{i})= \left( \frac{1}{2}\partial H^{+} \, , \; \ e^{\pm i \sqrt{2}H^{+}}
 \right)
$$
where,
$$
{\bf {G}} = \frac{G_{1}+iG_{2}}{\sqrt{2}} \, , \;\;\;\;\;\;\;\;   {\bf {\tilde
 {G}}}= \frac{G_{4}+iG_{3}}{\sqrt{2}}
\, .
\eqno(9)
$$

The above expressions show clearly that ${\bf {G}}$ and ${\bf {\tilde {G}}}$
 are both doublets under $ SU(2)_{H^{+}}$. The $ H^{+}$ factorization in the
 supercurrents it is not a particular property of the free-field realization
 it is a generic property for any $ {\bf{\hat{c}}}=4$ system with $N=4$
symmetry
 since it follows from the the$ N=4$ superconformal symmetry \cite{bd}. As a
 consequence the supercurrents  always have a factorized  product form in terms
 of two kinds of conformal operators; the ones do not have any dependence on
the
 $H^{+}$ field and they have conformal dimension $\frac{5}{4}$; the others are
 given uniquely in terms of $H^{+}$ and have conformal dimension $\frac{1}{4}$
 [see the expressions for ${\bf{G}}$ and ${\bf \tilde{G}}$ in eq. (9)].

We will now show  that it is possible to replace the four free
super-coordinates
 by those of the $\Delta_{k}^{(4)}$ curved space, keeping, however, the $N=4$
 symmetry in the deformed system.

\hfill\eject
\noindent
{\bf 2.2.  $\Delta_{k}^{(4)}$ realization. Parafermionic super-coordinates}

In the $\Delta_{k}^{(4)}$ space the coordinate currents are replaced by the
$\left(\frac{SU(2)}{U(1)}\right)_{k}$  compact \cite{ss} and
 $\left(\frac{SL(2,R)}{U(1)}\right)_{k'}$
non-compact \cite{tt},\cite{uu},\cite{mm} parafermions $\chi^{k}_{l}(\xi)$ and
$\chi^{k'} _{l}(\xi)$.

In both cases the parafermion algebra is defined by a collection of (non-local)
 currents $ \chi_{\pm l}(\xi)$, $l=0,1,2,... $ , [with $ \chi_{l}^{t}(\xi)=
 \chi_{-l}(\xi)$ and $ \chi_{0}^{t}(\xi)=\chi_{0}(\xi)=1$]. The parafermion
 currents satisfy the following OPE relations \cite{ss}--\cite{uu},
$$
\chi_{l_{1}}(\xi)\chi_{l_{2}}(\xi')\equiv
 C^{N}_{l_{1},l_{2}}(\xi-\xi')^{\Delta_{l_{1}+l{2}}-\Delta_{l_{1}}-
 \Delta_{l_{2}}} \chi_{l_{1}+l_{2}}~,
$$
$$
\chi_{l}(\xi)\chi_{l}^{t}(\xi') \equiv
(\xi-\xi')^{ -2 (\Delta_{l}-1)}\left(\frac{1}{(\xi-\xi')^2}+  \frac{2
 \Delta_{l}}{c(N)}  T_{\chi}(\xi') \right)
\, ,
\eqno(10)
$$
where $ T_{\chi}$ is the stress tensor of the parafermion theory with central
 charge $ c(N)=[3N/(N+2)-1]$
$$
T_{\chi}(\xi)T_{\chi}(\xi')\equiv \frac{c(N)}{2(\xi-\xi')^{4}}+
 \frac{2T_{\chi}(\xi')}{(\xi-\xi')^2}+ \frac{\partial
 T_{\chi}(\xi')}{(\xi-\xi')}
\, .
\eqno(11)
$$
 The $ \Delta_{l}$ are the conformal dimensions of $\chi_{l}$ fields, which are
 given by
$$
\Delta_{l}= \frac{l(N-l)}{N}
\, .
\eqno(12)
$$
The $ C^{N}_{l_{1},l_{2}}$ are known structure constants, which are determined
by associativity; the conformal dimensions of the fields $\Delta_{l}$
are constrained to satisfy the recursion relation:
$$
\Delta_{l+1}+\Delta_{l-1}+2\Delta_{l}-\Delta_{2}+2\Delta_{1}=n_{l}
\, ,\,\,\,\,\,\, n_{l}=0,1,2...
\, .
\eqno(13)
$$

In the case of $ \left(\frac{SU(2)}{U(1)}\right)_{k}$ compact parafermions, the
 parameter $N$ is positive and  equal to the level $k~~ (N=k)$. In this case
 the number of    $\chi^{k}_{l}$ fields is restricted to be $finite$ [$l:0
 \leq l \leq(k-1)$], and $ C^{k}_{l_{1},l_{2}} $ are given by
$$
C^{k}_{l_{1},l_{2}} = \left[
 \frac{\Gamma(k-l_{1}+1)\Gamma(k-l_{2}+1)
 \Gamma(l_{1}+l_{2})}{\Gamma(l_{1}+l)
   \Gamma(l_{2}+1)\Gamma(k+1)\Gamma(k-l_{1}-l_{2})}\right]^{\frac{1}{2}}
\, .
\eqno(14)
$$

In the non-compact version $ \left(
\frac{SL(2,R)}{U(1)} \right)_{k'}$, the
 parameter $N$ is negative and opposite to the level $k'~~ (N=-k')$. Also, the
 number of  parafermion fields is now
 $infinite$~~ ($l:~0~\leq~l~\leq
 \infty$). The structure constants $C^{-k'}_{l_{1}, l_{2}} $ are also given by
 eq. (14) via the analytic continuation $k \rightarrow -k'$; one then obtains
$$
C^{-k'}_{l_{1},l_{2}}= \left[  \frac{ \Gamma(k'+l_{1}+l_{2}) \Gamma(k')
 \Gamma(l_{1}+l_{2}+1)}{ \Gamma(l_{1}+1) \Gamma(l_{2}+1) \Gamma(k'+l_{1})
 \Gamma(k'+l_{2})}  \right]^{\frac{1}{2}}
\, .
\eqno(15)
$$

 In both compact and non-compact cases, their unitary representations and their
 characters are well known \cite{kp}, \cite{bbk}. For our purpose, we need to
 identify the super-coordinate currents of the  $\Delta_{k}^{(4)}$ space in
 terms of the parafermion fields  $\chi^{k}_{l}$,  $\chi^{k'}_{l}$, and in
terms
 of $H^{+}$ and $H^{-}$. As  in the free-field realization, the world-sheet
 fermionic super-partners will still be expressed in terms of $H^{+}$ and
$H^{-}$, but now with deformed radius (see below).

 The coordinate current identification with the parafermions can  easily be
 obtained in the large $k$ and $k'$ limit. Indeed in this limit the modified
 (non-) local coordinate currents $P_{k}$ and $\Pi_{k'}$ must approach their
fla
 expressions given by   eq.~(6), and so they must have conformal dimensions
 almost equal to one. It is then clear that up to a rescaling $P_{k}$ and
 $\Pi_{k'}$ have to be identified with the compact $\chi^{k}_{l=1}$ and
non-compact $\chi^{k'}_{l=1}$ parafermions
$$
\left(\frac{SU(2)}{U(1)}\right)_{k}: \,\,\,\,\,\,\,\,\,\,\,
 P(\xi) \longrightarrow P_{k}(\xi) \equiv
 \sqrt{\frac{2k}{k+2}}\chi^{k}_{l=1}(\xi),
$$
$$
\left( \frac{SL(2,R)}{U(1)} \right)_{k'}: \,\,\,\,\,\,\,\,\,
 \Pi(\xi) \longrightarrow \Pi_{k'}(\xi) \equiv
 \sqrt{\frac{2k'}{k'-2}}\chi^{k'}_{l=1}(\xi)
\, .
\eqno(16)
$$

 From eq. (12) the conformal dimensions of the $\Delta_{k}^{(4)}$ coordinate
 currents $P_{k}$ and $\Pi_{k'}$ are equal to $h_{P_{k}}=1-\frac{1}{k}$ and
$h_{\Pi_{k'}}=1+\frac{1}{k'}=1+\frac{1}{k+2}$. Owing to the deviation from the
 free-field dimensionality ($h_{P}=h_{\Pi}=1$) the OPE relations among the
 deformed coordinate currents $P_{k}$ and $\Pi_{k'}$ are those given by the
 parafermion algebra (see eqs.~(10) for $l=1$ and the above definitions of
 $P_{k}$ and $\Pi_{k'}$):
$$
P_{k}(\xi)P_{k}(\xi')^{t} \equiv
 \left[\frac{k}{k+2}\frac{2}{(\xi-\xi')^2}+2T_{P_{k}}(\xi')\right]
 (\xi-\xi')^{\frac{2}{k}},
$$
$$
\Pi_{k'}(\xi)\Pi_{k'}(\xi')^{t} \equiv
 \left[\frac{k'}{k'-2}\frac{2}{(\xi-\xi')^2}+2T_{\Pi_{k'}}(\xi')\right]
 (\xi-\xi')^{-\frac{2}{k'}}
\, .
\eqno(17)
$$
Equations (17) generalize the free-current OPE relations of eq. (6) to those of
the
 $\Delta^{(4)}_{k}$ curved space.

As we will see below, the anomalous dimensionality of the parafermionic
 coordinate currents $P_{k}$ and $\Pi_{k'}$ can be consistently compensated by
 a suitable modification of their world-sheet fermionic super-partners. Since
 this modification has to be consistent with the $ N=4$, ${\bf {\hat{c}}}=4$
 algebra, the $H^{+}$ part is fixed and is identical to that of the free-field
 realization. The only  possible modification that remains consist of deforming
 the radius of the $H^{-}$ field by some $\frac{1}{k}$ corrections. The exact
 value of the deformed  $H^{-}$ radius is found to be
$$
R_{H^{-}}=\alpha \sqrt{2}, \,\,\,\, {\rm with}  \,\,\,\,\,\,\,\,\,\,
 \alpha=\sqrt{\frac{k+4}{k}}
\, .
\eqno(18)
$$
 The above choice of $R_{H^{-}}$ gives rise to four operators with conformal
 dimension almost equal to $\frac{1}{2}$ in the large $k$ and $k'$ limit, which
 are nothing but the deformed world-sheet super-partners of $P_{k}$ and
 $\Pi_{k'}$ in $\Delta_{k}^{(4)}$ space,
$$
\Psi_{P}\longrightarrow e^{i \frac{1}{\sqrt{2}}(-\alpha H^{-}+H^{+})},
 \,\,\,\,\,\,\,\,\,\,\,{\rm with} \,\,\,\,\,\,\,\,\,\,
 h_{\Psi_{P}}=\frac{1}{2}+\frac{1}{k},
$$
$$
\Psi_{\Pi}\longrightarrow e^{i \frac{1}{\sqrt{2}}
(+ \frac{1}{\alpha}H^{-}+ H{+})}, \,\,\,\,\,\,\,\,\,\,\, {\rm with}
 \,\,\,\,\,\,\,\,\,\, h_{\Psi_{\Pi}}=\frac{1}{2}-\frac{1}{k+4}
\, .
\eqno(19)
$$

Observe that in the limit, $k\longrightarrow \infty $, $\Psi_{P}$ and
 $\Psi_{\Pi}$ become the two complex world-sheet fermions with canonical
 dimensionality $h_{\Psi_{P}}=h_{\Psi_{\Pi}}=\frac{1}{2}$.
Although in the $\Delta^{(4)}_{k}$ space we have anomalous conformal weights
for
 the coordinate currents $P_{k},\Pi_{k'}$ as well as for their super-partners
 $\Psi_{P}$,$\Psi_{\Pi}$, we may define without any obstruction the $N=4$ basic
 operators as follows:
$$
T_{B}= - \frac{1}{2} \left((\partial H^{+})^{2}+(\partial H^{-})^{2}
 \right)+T_{P_{k}}+T_{\Pi_{k'}}
$$
$$
{\bf {G}} = -\left( \Pi^{t}_{k'} e^{- \frac{i}{\alpha \sqrt{2}}H^{-}} +
 P^{t}_{k} e^{+ \frac{i\alpha}{ \sqrt{2}} H^{-}} \right) e^{+ \frac{i}{
 \sqrt{2}}H^{+}}
$$
$$
{\bf {\tilde {G}}}= \left( \Pi_{k'} \;\; e^{+ \frac{i}{\alpha \sqrt{2}}H^{-}} -
    P_{k} \;\; e^{- \frac{i \alpha }{ \sqrt{2}} H^{-}} \right) e^{+ \frac{i}{
 \sqrt{2}}H^{+}}
$$
$$
(S_{i})= \left( \frac{1}{2}\partial H^{+} \, , \; \ e^{\pm i \sqrt{2}H^{+}}
 \right)
\, .
\eqno(20)
$$

The validity of the $N=4$ superconformal algebra follows from the OPE relations
 among $P_{k},P_{k}^{t},\Pi_{k'},\Pi_{k'}^{t},$ given in eq. (17), and those of
 the free fields $H^{+},H^{-}$.
\vskip 0.3cm
The existence of the $N=4$ symmetry in the $\Delta^{(4)}_{k}$ non-trivial space
is of main interest, since it gives us the possibility to construct a new class
of $exact$ and $stable$ string solutions around non-trivial gravitational and
 dilaton backgrounds. More explicitly, we can arrange the degrees of freedom of
 the ten supercoordinates in three superconformal
 systems \cite{bb}--\cite{ee}:
$$
\hat{c}({\rm total})=10=[\hat{c}=2]_{0}+[\hat{c}=4]_{1}+[\hat{c}=4]_{2}
\, .
\eqno(21)
$$

The $\hat{c}=2$ subsystem is saturated by two free superfields; in one
 variation of our solution, one of the two superfields is chosen to be the
 time-like supercoordinate and the other to be  one of the nine space-like
 supercoordinates. In other variations, the two supercoordinates are Euclidean
o
 even compactified on a one- or two-dimensional torus.

The remaining eight supercoordinates appear in groups of four in
 $[\hat{c}=4]_{1}$ and $[\hat{c}=4]_{2}$. Both $[\hat{c}=4]_{A},\,\,\,\,A=1,2$,
 subsystems show an $N=4$ superconformal symmetry. The non-triviality of our
 solutions follows from the fact that there exist some realizations of the
$\hat{c}=4, N=4$
 superconformal systems,  which are based on geometrical and topological
non-trivial spaces other than the $T^{(4)}/Z_{2}$ orbifold and the $K_{3}$
 Calabi-Yau space. The $\Delta^{(4)}_{k}$ realization I presented above is one
 new example of such a system.

 There is another known non-trivial realization that shares the same
 superconformal symmetries, namely that of the $semi$-$wormhole$
 $four$-$dimensional$ $space$, $W^{(4)}_{k}$,\cite{hh},\cite{bb}--\cite{kk}.
For completeness, I will present below the basic operators and fields of the
 $W_k^{(4)}$ realization \cite{ggg},\cite{hh} as well as the additional
 realization based on $C^{(4)}_k$. It turns out that the $C^{(4)}_{k}$ space
is
 the dual of $W_k^{(4)}$ \cite{wd},\cite{kkl} and that it is also related to
som
 analytic continuations of  previous constructions \cite{qq},\cite{rr}.

\vfill\eject
\noindent
{\bf 2.3. $W_k^{(4)}$ semi-wormhole realization. $SU(2)_k\otimes U(1)_Q$
supercoordinates} \cite{ggg},\cite{hh}

The realization $W_k^{(4)}$ is based on a supersymmetric $SU(2)_k\times U(1)_Q$
WZW model, with a background term $Q = \sqrt{\frac{2}{k+2}}$ in the $U(1)_Q$
current. The four fermions of the model are free and are parametrized by
the $H^+$ and $H^-$ fields (via bosonization) as in the free-field
realization [eqs. (7), (8)]. The four-coordinate currents are the three
$SU(2)_k~~ (J^i, i = 1,2,3)$ currents and  one of the $U(1)_Q ~~(J_4)$
currents:
$$
J^i(\xi )J^j(\xi ')\equiv -\frac{k}{2}~\frac{\delta^{ij}}{(\xi -\xi ')^2} +
\epsilon^{ij\ell}~\frac{J^{\ell}}{(\xi -\xi ')^2}~,~~i = 1,2,3;
$$
$$
J^4(\xi )J^4(\xi ') \equiv \frac{-1}{Q^2(\xi -\xi ')^2}
\eqno(22)
$$
(the $Q$ rescaling of $J^4$ is for convenience).

The $T_B, {\bf {G, \tilde G}}$ and $S_i$ associated to $W_k^{(4)}$ are:
$$
T_{B} = -\frac{1}{2}~\bigg[ (\partial H^+)^2 + (\partial H^-)^2 + Q^2~
(J^2_1+J^2_2+J^2_3+J^2_4+\partial J_4)\bigg]
$$
$$
{\bf G } = Q~\bigg[(J_4 -i(J_3+\sqrt{2}\partial H^-))~
 e^{-i\frac{1}{\sqrt{2}}H^-}+(J_1-iJ_2)~e^{+i\frac{1}{\sqrt{2}}H^{-}} \bigg]
{}~e^{i\frac{1}{\sqrt{2}}H^+}
$$
$$
{\bf \tilde G} = Q~\bigg[(J_4 + i(J_3+\sqrt{2}\partial H^-))~
 e^{+i\frac{1}{\sqrt{2}}H^-}-(J_1+iJ_2)~
 e^{-i\frac{1}{\sqrt{2}}H^{-}}\bigg]
{}~e^{i\frac{1}{\sqrt{2}}H^+}
$$
$$
S_{0} = \frac{1}{\sqrt{2}} \partial H^+~,~~S_{\pm} = e^{\pm i\sqrt{2}H^+}
\, .
\eqno(23)
$$

In the above expressions the level $k$ of the $SU(2)_{k}$ and the background
 charge $Q$ are related because of the $N=4$ symmetry. Because of this
relation,
the
 central charge $\hat{c}(W_k^{(4)})=4$ for any value of the level $k$
 \cite{ggg}.
$$
\hat c[SU(2)_k] = \frac{2}{3} [3-\frac{6}{k+2} + \frac{3}{2}] = 3 -
\frac{4}{k+2},
$$
$$
\hat c[U(1)_Q] = \frac{2}{3} [1 + 3Q^2 + \frac{1}{2}] = 1 +2Q^2
$$
(the contributions $\frac{3}{2}$ and
$\frac{1}{2}$ inside $[\ldots ]$ in the
first and second line are those of the 3+1 free fermions)
$$
\hat c[SU(2)_k] + \hat c[U(1)_Q] = 4 + 2(Q^2 - \frac{2}{k+2})~,
$$
and so $\hat c [ W_k^{(4)}] = 4$ only if
$$
Q = \sqrt{\frac{2}{k+2}}
\, .
\eqno(24)
$$

The existence of $N = 4$, $\hat c = 4$ superconformal symmetry with this
value of $Q$ is found in ref. \cite{ggg}. What is extremely interesting is
the background interpretation of the $W_k^{(4)}$ space in terms of a
four-dimensional (semi-) wormhole space given by Callan, Harvey and
Strominger in ref. \cite{hh}. Indeed, for large $k$, the three $SU(2)_k$
coordinates define a three-dimensional subspace with a non-trivial topology
$S^3$, while the fourth coordinate with a background charge corresponds to
the scale factor of the $S^3$ sphere.
In the $W_k^{(4)}$ realization [see eq. (23)], both $H^+$ and $H^-$ are
 compactified on a torus with radius $R_{H^+} = R_{H^-} = \sqrt{2}$ as in the
 free-field case. There are in total three underlying $SU(2)$ Kac-Moody
 currents:
\begin{description}\item[~~(i)] the $SU(2)_k$ defined by the coordinate
currents
\item[~(ii)] the $SU(2)^+_1$ defined by the $H^+$ field with $R_{H^+} =
\sqrt{2}????????$
\item[(iii)] the $SU(2)^-_1$ defined by the $H^-$ field  with $R_{H^-} =
\sqrt{2}????????$
\end{description}

The background term in $T_B, Q\partial J^4$, comes from a non-trivial
dilaton background: $\Phi = Q X^4$.

The term $Q\sqrt{2}\partial H^-~\exp[\pm i\frac{1}{\sqrt{2}}H^- +
 i\frac{1}{\sqrt{2}}H^+]$ in ${\bf G}$ and ${\bf \tilde G}$ describes, at the
 same time, the standard fermionic torsion term $\pm Q\psi^i\psi^j\psi^k ~~(i =
 1,2,3,4)$, as well as the fermionic background term $\pm Q\partial\psi^i$.
 The torsion and background terms in the supercurrents are arranged together in
 $\sqrt{2}\partial H^-$, which appears as a shift of the $J_3$ current.

As we will see later on, the geometrical as well as the topological structure
of
 the $\Delta_k^{(4)}$ space differs from that of the $W_k^{(4)}$. In
$W_k^{(4)}$
 there is a non-vanishing torsion due to the WZW term, which is proportional to
 the $SU(2)$ structure constant. So, in $W_k^{(4)}$ there is a non-vanishing
 antisymmetric field background $B_{ij}$ with non-trivial field strength
 $H_{ijk}$,
$$
H_{ijk} = e^{-2\phi}\epsilon_{ijk}~{^{\ell}\partial_{\ell}}\Phi \simeq
e^{-2\phi}Q \epsilon_{ijk}^{\phantom{ijk}4}
\, .
\eqno(25)
$$

In order to be more explicit in our comparison with the $\Delta_k^{(4)}$ and
 $C^{(4)}_{k}$ spaces,  we give below the metric and the dilaton function of
the
 $W_k^{(4)}$ space in terms of two complex coordinates $z$ and $w$,
$$
ds\left[W_k^{(4)}\right]=k \frac{dzd\bar{z}+dwd\bar{w}}{z\bar{z}+w\bar{w}}
$$
$$
2\Phi=\log(z\bar{z}+w\bar{w})
\, .
\eqno(26)
$$
In $\Delta_k^{(4)}$ and $C^{(4)}_{k}$ the torsion is zero, while  the dilaton
 functions are non trivial (see below).
\vskip 1cm

\noindent
{\bf 3. THE TARGET SPACE INTERPRETATION OF THE $\Delta_k^{(4)}$ SPACE AND THE
$C^{(4)}_K$ REALIZATION}

\vspace*{0.3cm}
\noindent
{\bf 3.1. The semiclassical limit for $\Delta^{(4)}_k$ space}

For large $k$, the $\Delta_k^{(4)}$ space has a dimensional target-space
interpretation based on a non-trivial background metric $G_{ij}$ and on the
dilaton field $\Phi$ given by
 \cite{lll}--\cite{ee},\cite{qq},\cite{rr}--\cite{pp}:
$$
ds^2 = k\bigg\{(d\alpha )^2 + \tan^2\alpha d\theta^2\bigg\} + k'
\bigg\{(d\beta )^2  + \tanh^2\beta d\varphi^2\bigg\}
$$
$$
2\Phi = \log \cos^2\alpha + \log \cosh^2\beta + {\rm const.}
\, ,
\eqno(27)
$$
with
$$
\alpha \in \bigg[0,\frac{\pi}{2}\bigg] \bigcup \bigg[\pi ,\frac{3}{2}\pi
\bigg]~,~~\beta\in [0,\infty ]~,~~\theta , \varphi \in [0,2\pi].
$$
The term proportional to $k$ in eq. (26) parametrizes the two-dimensional
subspace defined by the $\bigg [\frac{SU(2)}{U(1)}\bigg]_k$ parafermionic
theory and the term proportional to $k'$ is the two-dimensional
subspace that is defined by the non-compact parafermionic theory based on the
$\bigg[\frac{SL(2,R)}{U(1)}\bigg]_{k'}$ axial gauged WZW model. It is well
known that a different metric $\tilde G_{ij}$ and dilaton function
$\tilde\Phi$ are obtained if one chooses a vector gauging instead of the
axial one \cite{pp},\cite{nn}
$(U(1)^{V}\otimes U(1)^{V})$:

$$
d\tilde {s}^2 = ~~k\bigg\{  (d\alpha )^2 + \frac{1}{\tan^2\alpha}
d\theta^2\bigg\}  + k'\bigg\{ (d\beta
)^2 + \frac{1}{\tanh^2\beta} d\varphi^2\bigg\}
$$
$$
2\tilde\Phi = ~~\log\sin^2\alpha
+ \log\sinh^2\beta +  {\rm const.}
\, .
\eqno(28)
$$

In both versions of gauging, $\Delta_k^{(4)}$ has always one non-compact
coordinate $(\beta )$ and three compact ones $(\alpha ,\theta ,\varphi )$;
$(G_{ij},\Phi )$ and $(\tilde G_{ij},\tilde\Phi )$ are related by a
generalized duality transformation \cite{oo}:
$$
R(t)\rightarrow \frac{1}{R(t)}~,~~\Phi\rightarrow\Phi + \log R(t)
\, .
\eqno(29)
$$
For later purposes, it is more convenient to use the complex notation:
$$
z = (\sin\alpha )e^{i\theta}~, ~~\omega = (\sinh\beta )e^{i\varphi}~,~~
(\rm axial~case)
$$
$$
\tilde z = (\cos\alpha )e^{i\theta}~,~~ y = (\cosh\beta )e^{i\varphi}~,
{}~~(\rm vector~case)
\, .
\eqno(30)
$$
In terms of $z$ and $w$, $G_{ij}$ and $\Phi$ are given by
$$
(ds)^2 = k~\frac{dzd\bar z}{1-z\bar z} + k'~\frac{dwd\bar w}{w\bar w + 1}~,
$$
and
$$
2\Phi = \log (1-z\bar z ) + \log (w\bar w + 1) + {\rm const.}
\, .
\eqno(31)
$$
The dual metrics $\tilde G_{ij}$ and $\tilde\Phi$ are given in terms of
$x$ and $y$ as:
$$
(d\tilde s)^2 = k~\frac{dxd\bar x}{1-x\bar x} +
k'~\frac{dyd\bar y}{y\bar y - 1}~,
$$
$$
2\tilde\Phi  = \log (1-x\bar x ) + \log (y\bar y - 1)~+ {\rm const.}
\, .
\eqno(32)
$$

 From eqs. (31) and (32), one observes that the
$\bigg(\frac{SU(2)}{U(1)}\bigg)_k$
 metric and dilaton function are self-dual
 ($z$ and $x$) while the $\bigg(\frac{SU(2,R)}{U(1)}\bigg)_{k'}$ are not.
The $w$ subspace is regular while the $y$ subspace is singular.
The $z$ subspace (or $x$) defines a $two$-$dimensional$ $bell$. Its
metric $G_{z\bar z}$, the Ricci tensor $R_{z\bar z}$, and its scalar
curvature $R^{(z)}$ are singular at the boundary of the bell $(z = 1)$:
$$
G_{z\bar z} = \frac{k}{1-z\bar z}~,~~R_{z\bar z} = \frac{-1}{(1-z\bar
z)^2~}, ~~R^{(z)} =  \frac{-1}{k(1-z\bar z)}~
\, .
\eqno(33)
$$
The $z$-bell is negatively curved for any value of $\vert z\vert < 1$.
\vskip 0.3cm
The $w$ subspace is regular everywhere, with finite positive curvature, and
is asymptotically flat for $\vert w\vert \rightarrow\infty$:
$$
G_{w\bar w} = \frac{k'}{w\bar w+1}~,~~R_{w\bar w} = \frac{+1}{(w\bar
w+1)^2}~, ~~R^{(w)} =  \frac{+1}{k'(w\bar w+1)}~
\, .
\eqno(34)
$$
It has a $cigar$ shape, with maximal curvature at $w = 0$.
\vskip 0.3cm
The $y$ subspace has a  different shape from its $w$-dual. It looks like a
$two$-$dimensional$ $trumpet$ with infinite curvature at the boundary $(y =
1)$.
It is  positively curved everywhere $(\vert y\vert >1)$ and is
asymptotically flat for $\vert y\vert\rightarrow\infty$:
$$
G_{y\bar y} = \frac{k'}{y\bar y-1}~,~~R_{y\bar y} = \frac{+1}{(y\bar
y-1)^2}~, ~~R^{(y)} =  \frac{+1}{k'(y\bar y-1)}
\, .
\eqno(35)
$$
We have therefore two different versions of the $\Delta_k^{(4)}$ space.
The first version is  the $(z,w)$-{\bf bell-cigar} four-dimensional space
and the second the $(x,y)$-{\bf bell-trumpet} one.

Up to now we have discussed the geometrical structure of the bosonic
coordinates of the $\Delta_k^{(4)}$ space. In order to complete the
 $\sigma$-model description of our solution, we must include the fermionic
 superpartners of $(z,w)$ coordinates, e.g. the four Weyl-Majorana left-handed
  $(\psi^a_+, a = 1,2,3,4)$, as well as the four Weyl-Majorana right-handed
 $(\psi^a_-, a = 1,2,3,4)$ ones.
Because of the $N = 1$ local supersymmetry, the interactions among
fermions are fixed in terms of the two-dimensional $\sigma$-model
backgrounds $G_{ij}, B_{ij}$ and $\Phi$. In the superconformal gauge, one
has the following generic form for the $N = 1$ $\sigma$-model action ($B_{ij} =
 0$ in $\Delta_k^{(4)}$ space):
$$
S = -\frac{1}{2\pi} \int d\xi d\bar\xi \bigg\{ V_+^aV_-^a - \frac{1}{2}
(\psi^a_+\nabla_-\psi^a_+ - \psi^a_-\nabla_+\psi^a_-)
-\frac{1}{2} R_{ab,cd} \psi^a_+\psi^b_+\psi^c_-\psi^d_- +\Phi R^{(2)}\bigg\}
\, ,
\eqno(36)
$$
where $a = 1,2,3,4$ are local flat indices and
$$
V_+^a = E^a_i\partial X^i~,~~V^a_- = E_i^a \bar\partial X^i~,~~{\rm
with}~~ G_{ij} = E_i^aE^a_j
\, .
\eqno(37)
$$
The $\nabla_-$ and $\nabla_+$  denote the left- and right-handed covariant
derivatives acting on left- and right-handed fermions $\psi^a_+ ,
\psi^a_-$:
$$
\nabla_-\psi^a_+ = \bar\partial\psi_+^a +
\Gamma^a_{\phantom{a}bc}\psi^c_+V^b_- ,
$$
$$
\nabla_+\psi^a_- = \partial\psi_-^a +
\Gamma^a_{\phantom{a}bc}\psi^c_-V^b_+
\, .
\eqno(38)
$$
Since $\Delta_k^{(4)}$ is defined as a direct product of two
two-dimensional subspaces, the $\sigma$-model action can be written as:
$$
S(\Delta_k^{(4)}) = S\bigg[\frac{SU(2)}{U(1)}\bigg]_k +
S\bigg[\frac{SL(2,R)}{U(1)}\bigg]_{k'}
\, .
\eqno(39)
$$
Because of the identity
$$
R_{ab,cd}\psi^a_+\psi^b_+\psi^c_-\psi^d_- = 2R
\psi^1_+\psi^2_+\psi^1_-\psi^2_-~
\, ,
\eqno(40)
$$
valid in both two-dimensional-subspaces, and thanks to the relations
$$
-(R^{(z)} + \Gamma_zG^{z\bar z}\Gamma_{\bar z}) = \frac{1}{k}~;~~
\bigg(\frac{SU(2)}{U(1)}\bigg)_k ,
$$
$$
-(R^{(w)} + \Gamma_wG^{w\bar w}\Gamma_{\bar w}) = \frac{-1}{k'}~;~~
\bigg(\frac{SL(2,R)}{U(1)}\bigg)_{k'}
\, ,
\eqno(41)
$$
it is then possible to rewrite the $\sigma$-model action in a more convenient
form,  which shows in particular (at least at the classical level) that
the fermions can be described in terms of two free bosonic fields
compactified in a special radius. Indeed, using eqs. (40) and (41), one
finds:
$$
S\bigg[\frac{SU(2)}{U(1)}\bigg]_k =
\frac{-1}{4\pi} \int d\xi d\bar\xi
\bigg\{ \partial A\bar\partial A + t(A)^2
\bigg(\partial\theta_A - \sqrt{\frac{2}{k}}~\psi^1_+\psi_+^2\bigg)~
\bigg(\bar\partial\theta_A - \sqrt{\frac{2}{k}}~\psi^1_-\psi^2_-\bigg)
$$
$$
-~(\psi^1_+\bar\partial\psi^2_+ + \psi^2_+\bar\partial\psi^2_+ +
\psi^1_-\partial\psi^1_- + \psi^2_-\partial\psi^2_-)
$$
$$
 +~\frac{2}{k} (\psi^1_+\psi^2_+)~(\psi^1_-\psi^2_-) + \log
C^2(A)R^{(2)}\bigg\}
\, ,
\eqno(42)
$$
where $A, \theta_A$ are rescaled fields so that, in the large $k$ limit, they
ar
 conventionally normalized as two real bosonic fields. In terms of $z$ and
 $\bar{z}$, the $A$ and $\theta_A$ are defined by the following relations:
$$
\sin^2~\frac{A}{\sqrt{2k}} = \sin^2\alpha = z\bar z
$$
$$
-i~\frac{1}{2}~\log~\frac{z}{\bar z} = \theta = \frac{\theta_A}{\sqrt{2k}}
$$
$$
t(A)^2 = \frac{z\bar z}{1-z\bar z} = \tan^2~\frac{A}{\sqrt{2k}}
$$
$$
C(A)^2 = 1-z\bar z = \cos^2~\frac{A}{\sqrt{2k}}
\, .
\eqno(43)
$$

In a similar way, the
$$
S\bigg[\frac{SL(2,R)}{U(1)}\bigg]_{k'} =
\frac{-1}{4\pi} \int d\xi d\bar\xi
\bigg\{ \partial  B\bar\partial B + T(B)^2
\bigg(\partial\theta_B - \sqrt{\frac{2}{k'}}~\psi^3_+\psi_+^4\bigg)~
\bigg(\bar\partial\theta_B - \sqrt{\frac{2}{k'}}~\psi^3_-\psi^4_-\bigg)
$$
$$
-~(\psi^3_+\bar\partial\psi^4_+ + \psi^4_+\bar\partial\psi^4_+ +
\psi^3_-\partial\psi^3_- + \psi^4_-\partial\psi^4_-)
$$
$$
 -~\frac{2}{k'} (\psi^3_+\psi^4_+)~(\psi^3_-\psi^4_-) + \log
C(B)^2R^{(2)}\bigg\}
\, ,
\eqno(44)
$$
where now $B$ and $\theta_B$ are defined in terms of $w$ fields:
$$
\sinh^2~\frac{B}{\sqrt{2k'}} = \sin^2\beta = w\bar w
$$
$$
-i~\frac{1}{2}~\log~\frac{w}{\bar w} = \varphi =
\frac{\theta_B}{\sqrt{2k}}
$$
$$
T(B)^2 = \frac{w\bar w}{1+w\bar w} = \tanh^2~\frac{B}{\sqrt{2k'}}
$$
$$
C(B)^2 = 1+w\bar w = \cosh^2~\frac{B}{\sqrt{2k'}}
\, .
\eqno(45)
$$

In both eqs. (42) and (44), the pure fermionic part of the action is given by
 the free-fermion kinetic terms together with some current-current
interaction. This fact permits us to describe the four left and the four
right fermions in terms of free bosonic fields $\phi_A, \phi_B$
compactified in the shifted radii:
$$
R^2_A = 1 + \frac{2}{k}  = \frac{k+2}{k}~;~~~~
\bigg(\frac{SU(2)}{U(1)}\bigg)_k ,
$$
$$
R^2_B = 1 - \frac{2}{k'}  = \frac{k'-2}{k'}~;~~
\bigg(\frac{SL(2,R)}{U(1)}\bigg)_{k'}
\, .
\eqno(46)
$$

The deviation from the value $R_A = R_B = 1$ is due to current-current
interactions. The decoupling of $\phi_A$ and $\phi_B$ fields can be seen
via the bosonization of fermions and the redefinition of the
$\partial\theta_A$ and $\partial\theta_B$ bosonic currents order by
 order in a $\frac{1}{k}$ or $\frac{1}{k'}$ expansion. This statement is indeed
 exact and it follows from the fact that both
$\bigg(\frac{SU(2)}{U(1)}\bigg)_k$
and  $\bigg(\frac{SL(2,R)}{U(1)}\bigg)_{k'}$ are exact $N = (2,2)$
superconformal models.
\vskip 0.3cm
This property is well known in both $\frac{SU(2)}{U(1)}$
and $\frac{SL(2,R)}{U(1)}$ supersymmetric-coset models in the compact and
non-compact parafermionic representations. The $N = 2$ generators, $J(\xi
), G(\xi ),  G^{t}(\xi )$ and $T(\xi )$, are given in terms of free-scalar
fields $(\phi_A,\phi_B)$ and in terms of (non-local) parafermionic currents
$(P_k,\Pi_{k'})$,\cite{ss}--\cite{vv}, [see also eqs.~(10), (16) and
 (17)].
\vskip 0.3cm
{}~(i) $\bigg(\frac{SU(2)}{U(1)}\bigg)_k$
$$
J(\xi ) = \sqrt{\frac{k}{k+2}}\partial\phi_A
$$
$$
G (\xi ) = P_{k}~e^{+i\sqrt{\frac{k+2}{k}}\phi_A}
$$
$$
G^{t} (\xi ) = P^{(t)}_{k}~e^{-i\sqrt{\frac{k+2}{k}}\phi_A}
$$
$$
T(\xi ) = -\frac{1}{2}(\partial\phi_A)^2 + T_{P_{k}}(\xi )
\, .
\eqno(47)
$$

{}~(ii)  $\bigg(\frac{SL(2,R)}{U(1)}\bigg)_{k'}$
$$
J(\xi ) = \sqrt{\frac{k'}{k'-2}}\partial\phi_B
$$
$$
G (\xi ) = \Pi_{k'}~e^{+i\sqrt{\frac{k'-2}{k'}}\phi_B}
$$
$$
G^{t} (\xi ) = \Pi^{(t)}_{k'}~e^{-i\sqrt{\frac{k'-2}{k'}}\phi_B}
$$
$$
 T(\xi ) = -\frac{1}{2}(\partial\phi_B)^2 + T_{\Pi_{k'}}(\xi )
\, ,
\eqno(48)
$$

$\phi_A$ and $\phi_B$ are free bosons compactified on radii $R_A =
\sqrt{\frac{k+2}{k}}$ and $R_B = \sqrt{\frac{k'-2}{k'}}$ and parametrize the
fermions $\psi^a_{\pm}$, $a = 1,2,3,4$, which appear in the $\sigma$-model
actions in eqs. (42) and (43) according to our previous discussion.

For large $k, k'$, the operators $e^{i\sqrt{\frac{k+2}{k}}\phi_A}$ and
$e^{i\sqrt{\frac{k'-2}{k'}}\phi_B}$  have conformal dimensions of almost
 $\frac{1}{2}~~(h_A = \frac{1}{2} + \frac{1}{k},~h_B = \frac{1}{2} -
 \frac{1}{k'}$);
 $P_{k}$ and $\Pi_{k'}$ are the parafermionic currents with dimensions $h_P = 1
 - \frac{1}{k}$ and $h_{\Pi} = 1 + \frac{1}{k'}$, so that $G(\xi )$ in eqs.
(47)
 and (48) are the $N = 2$ supercurrents of dimension $\frac{3}{2}$.

In order to show that for $k' = k+4$ the $N = 2$ superconformal symmetry is
 extended to $N = 4$ with $\hat c = 4 $, it is more convenient to work in terms
of
 the $H^{+}$ and $H^{-}$ fields defined in eqs. (7) and (8), instead of the
 $\phi_A$ and $\phi_B$ given above.
Both ($H^{+}$, $H^{-}$) and ($\phi_A$, $\phi_B$) are  bosonizations of  the
 same two-dimensional fermions and so they are related:
$$
\sqrt{\frac{k+2}{k}}\phi_A=\frac{1}{\sqrt{2}}(-\alpha H^{-}+ H^{+})
$$
$$
 \sqrt{\frac{k'-2}{k'}}\phi_B=\sqrt{\frac{k+2}{k+4}}\phi_B=
\frac{1}{\sqrt{2}}\bigg(+\frac{1}{\alpha}H^{-}+ H^{+}\bigg)
$$
with
$$
\alpha=\sqrt{\frac{k+4}{k}}
\, .
\eqno(49)
$$

In terms of the $H^+ ,H^-$ representation the $N = 2$ current of the
 $\Delta_k^{(4)}$ system is given uniquely in terms of the $\partial H^{+}$
$U(1)$
 current. The extension of the $N=2$ to an $N=4$ symmetry is due to:
\begin{description}
\item[~(i)] the extension of the $U(1)_{H^{+}}$ current algebra to an $SU(2)_1$
 self-dual compactification point with radius $R_{H^{+}}=\sqrt{2}$);
\item[(ii)] the existence of two (complex) supercurrents {\bf $G$} and
 {\bf{$\tilde{G}$}}  as they are given in  eqs. (20).
\end{description}

\vfill\eject
\noindent
 {\bf  3.2.  $C^{(4)}_{k}$ torus-bell realization.
 $\bigg(\frac{SU(2)}{U(1)}\bigg)_{k}\otimes U(1)_{R}\otimes U(1)_{Q}$
 supercoordinates}

The $C^{(4)}_{k}$ space is based on a supersymmetric gauged WZW model
$\bigg(\frac{SU(2)}{U(1)}\bigg)_{k}\otimes U(1)_{R}\otimes U(1)_{Q}$ with a
 background term
$Q=\sqrt{\frac{2}{k+2}}$ in  one of the two coordinate currents ($U(1)_{Q}$).
The other free  coordinate ($U(1)_{R}$) is compactified on a torus with radius
$R=\sqrt{2k}$.

In this realization the two coordinates $(x_1,x_2)$ are those of the
 $two-dimensional~bell$ space $(z, \bar z )$ of the
 $\bigg(\frac{SU(2)}{U(1)}\bigg)_{k}$ parafermions, $P_k$, $P^{t}_{k}$ [see
eq. (31)].
 The other two remaining coordinates $(x_3,x_4)$ are free, with a background
 charge $Q$ on $x_4$,
$$
J_{i}(\xi)J_{j}(\xi')\equiv \frac{-
 \delta_{ij}}{(\xi-\xi')^2}~~~~~~~{\rm with}~~~~~~i,j=3,4
\, .
\eqno(50)
$$

The fermions are parametrized in terms of the two bosons $\phi_A$ and $\phi_B$
 with radii
$R_A=\sqrt{\frac{k+2}{k}}$ and $R_B=1$.

The $N=2$ superconformal symmetry is manifest in the $C_k^{(4)}$ space since
bot
 sub-systems have $N=2$ superconformal symmetry. The $N=2$  generators of the
 ($P^t,P,\phi_A$) sub-system are given in eq. (47). The ($J_3,J_4,\phi_B)$
 Liouville-like  sub-system has the following $N=2$ generators \cite{abkl}:
$$
T_{Q}(\xi) = -\frac{1}{2}~\bigg[ (\partial \phi_B)^2 +J^2_3+J^2_4+ Q\partial
 J_4)\bigg]
$$
$$
 G (\xi) = \bigg[+J_4 +i(J_3 -Q\partial \phi_B)\bigg]~ e^{+i\phi_B}
$$
$$
 G^t(\xi) = \bigg[-J_4 + i(J_3 -Q\partial \phi_B)\bigg] ~e^{-i\phi_B}
$$
$$
J(\xi) = \partial \phi_B + QJ_3
\, .
\eqno(51)
$$

 In order to make  the $N=4$ symmetry manifest in the combined system, it is
 necessary to introduce the $H^{+}$  field, which defines the $SU(2)_1$
currents
 of the $N=4$ algebra. This can be done using the following field
redefinitions:
$$
\phi_B=\sqrt{\frac{1}{2}}(H^+ + H^-)
$$
$$
\phi_A=\sqrt{\frac{k}{k+2}}\bigg[~~\sqrt{\frac{1}{2}}(H^+ - H^-)
 -\sqrt{\frac{2}{k}}~~\partial \sigma \bigg]
$$
$$J^3 =\sqrt{\frac{2}{k+2}}\bigg[~~\sqrt{\frac{1}{2}}(H^+ - H^-) -
 \sqrt{\frac{k}{2}}~~\partial\sigma \bigg]
\, .
\eqno(52)$$

Using the above redefinisions the $N=4$ generators take the desired  factorized
 form in terms of $P^t,P,J^4,\partial\sigma,H^+$ and $H^-$.
$$
T_{B} = -\frac{1}{2}~\bigg[ (\partial H^+)^2 + (\partial H^-)^2 +
 (\partial\sigma)^2+J^2_4+ Q\partial J_4)\bigg] +T_{P_k}
$$
$$
{\bf G } = -\bigg[(-J_4
 +i(\sqrt{\frac{k}{k+2}}\partial\sigma-\sqrt{\frac{2}{k+2}}\partial H^-))~
 e^{-i\frac{1}{\sqrt{2}}H^-}+
P^t~ e^{+i\frac{1}{\sqrt{2}}H^{-}+i\sqrt{\frac{2}{k}}\sigma} \bigg]
{}~e^{i\frac{1}{\sqrt{2}}H^+}
$$
$$
{\bf \tilde G} =+\bigg[(+J_4
 +i(\sqrt{\frac{k}{k+2}}\partial\sigma-\sqrt{\frac{2}{k+2}}\partial H^-))~
e^{+i\frac{1}{\sqrt{2}}H^-}- P~
 e^{-i\frac{1}{\sqrt{2}}H^{-}-i\sqrt{\frac{2}{k}}\sigma} \bigg]
{}~e^{i\frac{1}{\sqrt{2}}H^+}
$$
$$
S_{0} = \frac{1}{\sqrt{2}} \partial H^+~,~~S_{\pm} = e^{\pm i\sqrt{2}H^+}
\, .
\eqno(53)
$$

The closure of the $N=4$ algebra can be easily verified using the OPE relations
of the $P^t,P$ parafermions in eq. (17) and those of the free bosonic fields
$H^+,H^-,x_4$
 and $\sigma$.
The $H^+$ and $H^-$ are compactified on radii $R_{H^+} = R_{H^-} = \sqrt{2}$,
 while the $\sigma$ is compactified with radius $R_{\sigma}=\sqrt{2k}$ (or its
 dual $\tilde{R_{\sigma}}=\sqrt{\frac{2}{k}}$).

For large $k$ the $C_k^{(4)}$ has a four-dimensional target-space
interpretation,
 which is a product of two sub-spaces; the first sub-space has the shape of a
 $two-dimensional~bell$ while the second sub-space is locally flat with the
shape
 of a $two-dimensional~cylinder$.
 The metric and the dilaton function on the $C_k^{(4)}$, {\bf torus-bell}
four-dimensional space are:
$$
ds(C_k^{(4)})= k\frac{dzd\bar z}{1-z\bar z } +k~dwd\bar w
$$
$$
2\Phi=w+\bar w +log(1-z\bar z)
\, .
\eqno(54)
$$

What is interesting is that the $C_k^{(4)}$ space can be obtained by performing
 (supersymmetric) duality transformation on $W^{(4)}_{k}$ \cite{wd},\cite{kkl}.
 On $C_k^{(4)}$ the torsion  is zero, while  it is non-trivial
on  $W^{(4)}_{k}$ [see
eq. (25)].
The  metric of $C_k^{(4)}$ is singular while that
  of $W^{(4)}_{k}$ is regular. The question about the relevance of the
 singularity in the ``stringy" constructed models is still an open question.
\vskip 1cm
\noindent
{\bf 4.  STRING CONSTRUCTIONS USING AS BUILDING BLOCKS
THE}\\  {\bf $F^{(4)}$,$\Delta^{(4)}_k$,$W^{(4)}_k$ AND $C^{(4)}_k$ N=4
SPACES}.

Having at our disposal non-trivial $N=4$, $\hat c =4$ superconformal systems,
 we can use them as building blocks to obtain  new classes of $exact$ and
 $stable$ string solutions around non-trivial backgrounds in both Type II and
 heterotic superstrings constructions.

 In the Type II case, we arrange the degrees of freedom of the ten
 supercoordinates in three superconformal systems:
$$
\hat c = 10 = \{\hat c = 2 \}_0 + \{\hat c = 4\}_1 + \{\hat c = 4\}_2~.
$$
The $\hat c=2$ system is saturated by two free superfields (compact or
non-compact), and so the background metric is flat:
$$
ds^2\{F^2\} = dxd\bar x~,~~(x = x_1+ix_2)~.
$$

The remaining eight supercoordinates appear in a group of four. Both
$\{\hat c = 4\}_{1,2}$ building blocks are  $N = 4$ superconformal systems
based
 on $F^{(4)}$, $W^{(4)}_k$, $C^{(4)}_k$ or on the two versions of
 $\Delta^{(4)}_k$ spaces. (The $F^{(4)}$ realization includes the compact or
 non-compact four-dimensional flat space as well as  the $T^4/Z_2$  orbifold
 models.)

 The advantage of this approach of constucting new solutions
compared with the subclass of models $(F^{(6)}\times W^{(4)})$ studied in the
literature (using the $\sigma$-model approach) \cite{hh}, lies in the complete
 knowledge of the $\{\hat c =2\}_0$ and  $\{\hat c = 4\}_A$ superconformal
 theories. With this knowledge, we are able to study the full string spectrum
 and derive the partition functions of the models and not only the background
 solutions for large $k_A$.

 The full spectrum around any constructed background is given in terms of
severa
 characters of known conformal theories:
\begin{description}
\item[~~(i)] in the $W^{(4)}_k$ realization \cite{bb}--\cite{dd}, one uses
modular invariant
 character combination of $SU(2)_k, SU(2)_{H^{+}}$ and $SU(2)_{H^{-}}$,
together
 with the $U(1)_Q$ Liouville-type characters;
\item[~(ii)] in the $\Delta^{(4)}_k$ realization, one uses the compact
\cite{kp} and
 non-compact \cite{bbk} parafermionic characters (string functions), together
 with the $SU(2)_{H^{+}}$ and $U(1)_{R_{H^-}}$ with \hfill\break $R_H =
\sqrt{2}~ \sqrt{\frac{k+4}{k}}$;
\item[(iii)] in the $C^{(4)}_k$ realization, one uses
the compact parafermionic characters, the
 $U(1)_Q$ Liouville-type characters, the two $SU(2)_{H^{+}}$ and
$SU(2)_{H^{-}}$
 as well as the $U(1)_{R_{\sigma}}$ characters (with $R_{\sigma}~=~\sqrt{2k}$);
\item[~(iv)] finally in $F^{(2)}$ and $F^{(4)}$, one uses  non-compact
flat-space characters,
 toroidal  ones or orbifold ones.
\end{description}

It is important to stress here that the above character combinations are not
arbitrary but  are dictated by the global existence of $N = 4$
superconformal symmetry as well as the modular invariance of the partition
 function.
These requirements define some generalized GSO projections
\cite{bb}--\cite{ee}, similar to that of the fermionic construction
\cite{ww} and that of the conformal block construction \cite{yy} of Gepner.
One of these projections is fundamental and guarantees the existence of
some space-time supersymmetries via the $N = 4$ spectral flows. It can be
expressed in terms of the two isospins of the $N = 4 ~~(\hat c = 4)_{A},~~
   A =1,2~~$ sub-systems ($ [SU(2)_{H^{+}}]_{A} $  $[j(s_1), j(s_2)]$)
 \cite{bb}--\cite{ee}:
$$
2(j(s_1) + j(s_2)) = {\rm odd~integer}~.
\eqno(55)
$$

Equation (55) guarantees the existence of some covariantized target-space
symmet
the proposed non-trivial backgrounds. This stabilizes our solutions (at least
perturbatively) and projects out all kinds of tachyonic or complex conformal
wei
states. This projection phenomenon is similar to the one observed by Kutasov
and
in the framework of non-critical superstrings with an $N = 2$ globally defined
superconformal symmetry \cite{zz}.

The stability of the vacuum under string-loop perturbation follows from the
 existence of a globally defined $N=4$ symmetry, which implies a new
realization
  target space supersymmetry around the non-trivial backgrounds, which are
 defined by the  $W^{(4)}_{k}$, $C^{(4)}_{k}$ and $\Delta^{(4)}_{k}$  spaces.
 This new target-space supersymmetry follows from  the $N=4$ spectral flow
 relations, which   imply a spectrum degeneracy among space-time bosonic and
 space-time  fermionic string excitations. This string spectrum degeneracy
 guarantees the vanishing of the partition function (at least at the one-string
 loop level), for all values of~$k$.

Heterotic solutions are simply obtained via a generalized
\cite{aai},\cite{yy} heterotic map. Here also the solutions are stable, but
the number of covariantized space-time supersymmetries is reduced by a
factor of two.

        We hope that our explicit construction of a family of consistent and
stable solutions will give a better understanding of some fundamental
string properties, especially in the case of strongly curved backgrounds
(small $k_A$), where the notion of  space-time dimensionality and
topology breaks down.
\vspace*{2cm}

\noindent
{\bf ACKNOWLEDGEMENTS}

I would like to thank  E. Kiritsis and D. L\"ust for the benefit I had during
our scientific discussions. I would like also to thank L. Alvarez-Gaum\'e, I.
Antoniadis, C. Callan, S. Ferrara and D. Gross for fruitful discussions.

\vfill\eject


\begin{thebibliography}{99}
\bibitem{all} R.C. Myers, \PL {\bf B199} (1987) 371;\\
K.A. Meissner and G. Veneziano, \PL {\bf B267} (1991) 33;\\
A. Sen, \PL {\bf B271} (1991) 295;\\
M. Gasperini, J. Maharana and G. Veneziano, \PL {\bf B272} (1991) 277;\\
M. Gasperini and G. Veneziano, \PL {\bf B277} (1992) 256;\\
A.A. Tseytlin, Class. Quant. Grav. {\bf 9} (1992) 979;\\
A.A. Tseytlin, Cambridge preprints DAMPT-92-15 and DAMPT-92-36;\\
I. Bars  and K. Sfetsos, Univ. Southern Calif. preprint USC-92/HEP-B1;\\
K. Behrndt, preprints DESY 92-055 and DESY 92-179;\\
M. Gasperini and G. Veneziano, preprint CERN-TH.6572/92;\\
H.J. de Vega and N. Sanchez, Univ. Paris VI preprint LPTHE 92-31;\\
H.J. de Vega, A.V. Mikhailov and N. Sanchez, Univ. Paris VI preprint LPTHE
92-32

\bibitem{aben} I. Antoniadis, C. Bachas, J. Ellis and D.V. Nanopoulos,
\PL {\bf B211} (1988) 393, \NP {\bf B328} (1989) 117;\\
I. Antoniadis, C. Bachas and A. Sagnotti, \PL {\bf B235} (1990) 255;\\
J. Polchinski, \NP {\bf B324} (1989) 123.

 \bibitem{hh} C. Callan, J. Harvey and A. Strominger, \NP {\bf B359}
(1991) 611;\\
C. Callan, Lectures at Sixth J.A. Swieca Summer School, Princeton preprint
PUPT-
(1991).


\bibitem{lll} E. Witten, \PR {\bf D44} (1991) 314.

\bibitem{mm} I. Bars, \NP {\bf B334} (1990) 125;\\
I. Bars and D. Nemeschansky, \NP {\bf B348} (1991) 89.

\bibitem{nn} R. Dijkgraaf, E. Verlinde and H. Verlinde, \NP {\bf B371}
(1992) 269.

\bibitem{bb} I. Antoniadis, S. Ferrara and C. Kounnas, LPTENS-CERN
preprint in preparation.

\bibitem{cc} I. Antoniadis and C. Kounnas, LPTENS-CERN preprint in
preparation.

\bibitem{dd} C. Kounnas, Proceedings of the  International Conference,
on ``Lepton-Photon Symposium and Europhysics Conference on High Energy
Physics", Geneva, 1991, Vol. 1, pp. 302-306.

\bibitem{ee} C. Kounnas, CERN preprint TH.6799/93.

\bibitem{jj} S.J. Rey, \PR {\bf D43} (1991) 526 and SLAC-PUB-5659 (1991).

\bibitem{kk} M. Bill\'o, P. Fr\'e, L. Girardello and A. Zaffaroni,
preprint SISSA 159/92/EP or IFUM/431/FT.

\bibitem{qq} P. Ginsparg and F. Quevedo, Nucl. Phys. {\bf B385} (1992) 527.
\bibitem{tsva} R. Brandenberger and C. Vafa, \NP {\bf B316} (1988) 391;\\
A.A. Tseytlin and C. Vafa, \NP {\bf B372} (1992) 443.




\bibitem{rr} C. Kounnas and D. L\"ust, \PL {\bf B289} (1992) 56.

\bibitem{lust} D. L\"ust, CERN preprint TH.6850/93 and Proccedings of ``4th
 Hellenic School on Elementary Particle Physics", Corfu,  1992.

\bibitem {nawit}C. Nappi and E. Witten, \PL {\bf B293} (1992) 309.

\bibitem {mul} M. Muller, \NP {\bf B337} (1990) 37.

\bibitem{hor} P. Horava, \PL {\bf B278} (1992) 101.

\bibitem{oo} A. Giveon, \MPL {\bf A6} (1991) 2843.

\bibitem{pp} E. Kiritsis, \MPL {\bf A6} (1991) 2871; CERN preprint TH.6797/93.




\bibitem{aaa} C. Callan, D. Friedman, E. Martinec and M. Perry, \NP {\bf
B262} (1985) 593;\\
E. Fradkin and A. Tseytlin, \NP {\bf B261} (1985) 1.

\bibitem{ff} M. Ademollo et al., \NP {\bf B114} (1976) 297;\\
T. Eguchi and A. Taormina, \PL {\bf B200} (1988) 634;\\
A. Sevrin, W. Troost and A. Van Proeyen, \PL {\bf B208} (1988) 447.

\bibitem{ggg} C. Kounnas, M. Porrati and B. Rostand, \PL {\bf B258} (1991)
61.

\bibitem{bd} T. Banks and L. Dixon, \NP {\bf B307} (1988) 93.




\bibitem{ss} A.B. Zamolodchikov and V.A. Fateev, \SPJ~ {\bf 62} (1985)
215.

\bibitem{tt} L. Dixon, J. Lykken and M. Peskin, \NP {\bf B325} (1989) 329.

\bibitem{uu} J. Lykken, \NP {\bf B313} (1989) 473.

\bibitem{vv} A.B. Zamolodchikov and V.A. Fateev, \SPJ ~{\bf 63} (1986) 913;\\
P. Di Vecchia, J. Petersen, M. Yu and H. Zheng, \PL {\bf B174} (1986) 280;\\
Z. Qiu, \PL {\bf B188} (1987) 207.

\bibitem{wd} M. Rocek, K. Schoutens and A. Sevrin, \PL {\bf B265} (1991)
303.

\bibitem{kkl} E. Kiritsis, C. Kounnas and D. L\"ust, CERN preprint in
preparatio


\bibitem{bkk} I. Bakas and E. Kiritsis, ENS preprint LPTENS-92-30 (1992).

\bibitem{bk} I. Bakas and E. Kiritsis, Nucl. Phys. {\bf B343} (1990) 185; Mod.
Phys. Lett.
{\bf A5} (1990) 2039.

\bibitem{kp} V. Kac and D. Peterson,  \AM {\bf 53} (1984) 125.

\bibitem{bbk} I. Bakas and E. Kiritsis,  \IJMP {\bf A7} [Suppl.
1A] (1992)
339.

\bibitem{abkl} I. Antoniadis, C. Bachas and C. Kounnas, \PL {\bf B243} (1990)
 185.


\bibitem{ww} I. Antoniadis, C. Bachas and C. Kounnas, \NP {\bf B289} (1987)
87;\\ H. Kawai, D.C. Lewellen and S.H.-H. Tye, \NP {\bf B288} (1987) 1.

\bibitem{yy} D. Gepner, \PL {\bf B199} (1987) 370; \NP {\bf B296} (1988)
757.

\bibitem{zz} D. Kutasov and N. Seiberg, \PL {\bf B251} (1990) 67.

\bibitem{aai} W. Lerche, D. L\"ust and A.N. Schellekens, \PL {\bf B181}
(1986) 71; \NP {\bf B287} (1987) 477.



\end{thebibliography}
\end{document}